\documentclass[a4paper,11pt]{article}

\usepackage{jcappub}

\usepackage{tensor}

\usepackage{xspace}

\newcommand{\Horava}{Ho\v{r}ava\xspace}

\def\beq{\begin{equation}}
\def\eeq{\end{equation}}

\newcommand{\lie}{\pounds}
\newcommand{\dd}{\mathrm{d}}
\newcommand{\eom}{\mathcal{E}}
\newcommand{\aeth}{\text{\ae}}
\newcommand{\cnstr}{\mathcal{C}}
\newcommand{\hc}{\mathcal{H}}

\newcommand{\gis}{\Sigma}   
\newcommand{\gih}{\Upsilon}  
\newcommand{\gix}{\Xi}   
\newcommand{\giv}{V}   
\newcommand{\gim}{M} 
\newcommand{\giz}{Z}  
\newcommand{\giw}{W} 
\newcommand{\eqcoord}{\overset{*}{=}}
\newcommand{\xs}{\beta} 
\newcommand{\xg}{\nu} 
\newcommand{\mm}{m} 
\newcommand{\gid}{{\delta^\text{F}}}

\def\vphi{\varphi}

\newcommand{\dm}{\text{dm}}

\begin{document}
\title{Ponderable aether}

\author[a,b]{Antony J. Speranza}
\affiliation[a]{Maryland Center for Fundamental Physics, University of Maryland, College Park, 
Maryland 20742}
\affiliation[b]{Perimeter Institute for Theoretical Physics, 31 Caroline Street North, ON N2L 2Y5, 
Canada}
\emailAdd{asperanz@umd.edu}

\date{April 13, 2015}
\abstract{
We consider a Lorentz-violating theory of gravity where the aether vector is 
taken to be nondynamical.  This ``ponderable aether theory'' is almost the 
same as Einstein-aether theory (where the aether vector is dynamical),
but involves  additional integration constants 
arising due to the loss
of initial value constraints. 
One of 
these   
produces an effective energy density for the aether fluid, similar to the 
appearance of dark matter in projectable \Horava gravity and the mimetic 
dark matter theory.
Here we investigate the extent to which this energy density can reproduce the 
phenomenology of dark matter.  
Although it is indistinguishable from cold dark
matter in homogeneous, isotropic cosmology, it encounters phenomenological problems 
in both spherically symmetric configurations 
and cosmological perturbations.  Furthermore, inflationary
considerations lead us to expect a tiny value for the ponderable aether energy density 
today unless a sourcing effect is added to the theory.  
The theory then effectively reduces to dynamical Einstein-aether theory, 
rendering moot the question of whether an aether must be 
dynamical in order to be consistent.  
}

\maketitle
\flushbottom

\section{Introduction}\label{sec:intro}

Some recent attempts to address  problems posed by cosmology and quantum gravity
involve imbuing spacetime with 
additional structure.  When such 
structure violates local Lorentz boost symmetry, it singles out a preferred time direction
at each point, defining an aether field $u^a$.  In a companion paper \cite{Jacobson2014}, 
we considered a 
variety of ways in which the aether can be constructed from  background structures,
i.e.\ fields that appear in the action but are fixed rather than varied in the action principle.  
Here, we analyze some of the phenomenological consequences of these theories in 
astrophysical and cosmological settings.

We will focus on two particular structures from ref.\ \cite{Jacobson2014}, the {\it fixed threading}
and the {\it fixed clock}.  A threading is a congruence of curves that foliate a manifold
\cite{Boersma1995, Boersma1995a}.  
In the fixed threading theory, the aether lies tangent to a threading that 
is fixed as background structure.  This fully determines
 the direction of the 
aether vector.  A clock is a scalar field $\psi$ that 
determines the overall scale of the aether vector via $u^a\nabla_a \psi =1$.  
When the aether is constrained to be a unit vector, 
$\psi$ corresponds to proper time along the flow lines  
of $u^a$, hence its interpretation as a ``clock."

 The aether theories  
constructed using
 these background structures resemble Einstein-aether theory, but with the crucial difference that some or all components of $u^a$ are nondynamical,
i.e.\ some of the aether equations of motion are not explicitly imposed.   In the ``least dynamical'' situation, with both a fixed threading and a fixed clock, none of the aether equations of motion are imposed, and the dynamics is solely determined through the Einstein equation.  Nevertheless, the Bianchi identity guarantees that 
 the 
 Lie derivative along $u^a$ of the Einstein-aether field equation vanishes.
 The theory is therefore equivalent to Einstein-aether theory
with a constant source  in the aether equations of motion.

The present work looks into the phenomenology of these sources in the aether field
equation.   The most notable effects come from {\it ponderable aethers}, which arise
in fixed clock theories.  The
corresponding source, $\tilde\mu$, functions as an energy density in the Einstein equation.   
In cosmological settings, it contributes 
a dust-like fluid component 
to the Friedmann equation.  This raises the question of whether it can account 
for the observed cold dark matter (CDM).  We investigate this idea on both small scales, 
involving the Newtonian limit and static spherically symmetric solutions, as well as  large 
scales, involving linear cosmological perturbations.   Both regimes possess significant 
phenomenological difficulties.  
There are no static, spherically symmetric solutions with 
nonzero values for the source densities,   
precluding a steady final state for spherically symmetric 
gravitational collapse.  What happens instead remains an open question. 
For cosmological perturbations, the source density $\tilde\mu$ 
leads to  rapidly growing scalar 
perturbations on superhorizon scales, which is at odds with observations of cosmological 
structure.  

In Ref.\ \cite{Jacobson2014}, we noted that the ponderable aether theory is closely
related other theories where dark matter arises as an integration variable.  Notably, when
the aether is constrained at the level of the action
to be geodesic and hypersurface-orthogonal, the result is the
IR limit of projectable \Horava gravity \cite{Horava2009}.  The ``dark matter as an 
integration constant''\cite{Mukohyama2009}
in that theory is analogous to the source density
$\tilde\mu$ for the ponderable aether theory.  Another similar theory with such a
dark matter component is the mimetic matter theory \cite{Chamseddine2013}, which 
was shown to be equivalent to Einstein gravity together with a constraint that 
the gradient $\nabla_a \vphi$ of a scalar field be a unit one-form \cite{Golovnev2014}.  
The arguments of Refs. \cite{Blas2009, Jacobson2014a} then show that the mimetic 
theory is equivalent to projectable \Horava gravity with the GR choice of 
coupling parameters.  
Since projectable \Horava gravity arises as the limit $c_\omega\rightarrow\infty$, 
$c_a\rightarrow\infty$ of Einstein-aether theory
\cite{Blas2010, Jacobson2013} ($c_\omega, c_a$ are defined in (\ref{eqn:K})), 
the results of this paper apply to the mimetic and projectable \Horava theories 
by taking this limit.  Note also that the vector mimetic theory 
of Ref.\ \cite{Barvinsky2014}
is closely related
to the ponderable aether, although the aether is dynamical in the former but 
nondynamical in the latter.

This work begins in section \ref{sec:theory} with an overview of  the theories  described in \cite{Jacobson2014},  and discusses  the dynamics when one or more of the aether equations are not imposed.  In particular, the constraints for the theory, described in \ref{sec:constraints}, require special attention.   In section \ref{sec:lvdm}, we show how the ponderable aether theory arises as a limit of  Lorentz-violating CDM \cite{Blas2012a}.  Following this, we proceed to phenomenology on small scales in section \ref{sec:smallpheno}, where the obstacle coming from static, spherical symmetry is discussed.  Finally, we explore cosmological solutions in section \ref{sec:cosmopheno}, focusing particularly on the challenges raised by the scalar perturbations.  Section \ref{sec:conclusion} provides a summary and discussion of the results.  

Throughout this paper we employ the metric signature $(+,-,-,-)$.  Latin letters denote abstract tensor indices, while Greek letters are used for coordinate indices.  Tildes are used to
represent tensor densities of weight +1.

\section{Lorentz-violating structures}\label{sec:theory}

\subsection{Dynamics for Lorentz-violation}\label{sec:action}
The coupling between $u^a$ and $g_{ab}$ is described through an action that incorporates all terms consistent with the  unbroken spacetime symmetries.  
When organized in a derivative expansion, the lowest order terms appear at two derivatives, and comprise the Einstein-aether action  \cite{Jacobson2001,Jacobson2008},
\begin{subequations}\label{eqn:aetheraction}
\begin{align}
S_{\ae}&=\frac{-1}{16\pi G_0} \int d^4x\sqrt{-g}\left(R+K\right), \label{eqn:aeaction}\\
K&=\frac{c_\theta}{3}\theta^2+c_\sigma \sigma^2+c_\omega \omega^2+c_a a^2. \label{eqn:K}
\end{align}
\end{subequations}
The terms  in (\ref{eqn:K}) are the expansion $\theta=\nabla_au^a$, acceleration $a_b = u^a\nabla_a u_b$, shear $\sigma_{ab} = \nabla_{(a}u_{b)}+\frac13\theta f_{ab} - u_{(a} a_{b)}$ ($f_{ab}\equiv u_a u_b-g_{ab}$ is the spatial metric) and twist $\omega_{ab} =  \nabla_{[a}u_{b]} - u_{[a} a_{b]}$ of the aether congruence.  The coupling constants appearing here are related to the usual constants $c_{1,2,3,4}$ by $c_\theta = c_1+3c_2+c_3$, $c_\sigma = c_1+c_3$, $c_\omega = c_1-c_3$, and $c_a = c_1+c_4$ \cite{Jacobson2013}.   
In addition, the unit constraint on $u^a$ can be imposed by adding a Lagrange multiplier term 
\begin{equation}
S_\lambda=-\int d^4x\sqrt{-g}\lambda\left(g_{ab}u^a u^b -1\right). \label{eqn:lagrange}
\end{equation}
Finally, the coupling to matter fields $\chi$ is described by a matter action $S_m[\chi, g_{ab}]$ which involves no direct interaction between $u^a$ and $\chi$.  The total action for the theory is then
\begin{equation}\label{eqn:fullaction}
S = S_{\ae}+S_\lambda + S_m.
\end{equation}

The various constructions of $u^a$ described in \cite{Jacobson2014} have the effect of producing theories where different components of $u^a$ can be considered nondynamical, i.e.\ parts of its equation of motion are not imposed.  Nevertheless, when the Einstein equation and matter field equations hold, the aether field equations are almost imposed; specifically, their derivative along  
the flow of $u^a$ vanishes,
\begin{equation} \label{eqn:lieaethereom}
\lie_u\left(\tilde\eom_a -2\tilde\lambda u_a\right) = 0. 
\end{equation}
Here the covector density $\tilde\eom_a \equiv  \dfrac{\delta S_{\ae}}{\delta u^a}$ is the functional derivative of the aether action (\ref{eqn:aeaction}) with respect to $u^a$ (see equation  (\ref{eqn:aethereom}) for its explicit form), and $\tilde\lambda = \sqrt{-g}\lambda$.  This relation follows from the diffeomorphism invariance of the action (\ref{eqn:fullaction}), and is equivalent to the conservation of the aether stress tensor.

These nondynamical aether theories can therefore be thought of as Einstein-aether theory supplemented with a source for the aether equation of motion, 
\begin{equation} \label{eqn:aethersource}
\tilde\eom_a -2\tilde\lambda u_a = -\tilde\mu_a,
\end{equation}
where the source must be invariant along 
the flow of $u^a$, 
\beq\label{liemutilde}
\lie_u \tilde\mu_a = 0.
\eeq
For different types of background structures, different components of $\tilde\mu_a$ are 
nonzero.  In the {\it fixed threading theory}, where the aether is constructed as tangent to
a nondynamical threading of spacetime by timelike curves, the quantity $\tilde\mu_a^\perp\equiv(\delta^b_a
-u^bu_a)\tilde\mu_b$ is nonzero. Alternatively, in the {\it fixed clock theory}, the parallel
component of the aether is determined by a  clock field background structure
$\psi$ by requiring $u\cdot\dd\psi = 1$.  In this theory, the quantity 
\beq
\tilde\mu\equiv \sqrt{-g}\mu \equiv u^a\tilde\mu_a
\eeq
 is nonzero.  The {\it fixed aether} theory possesses both a fixed 
threading
and a fixed clock, which together completely determine $u^a$.  
All components of $\tilde\mu_a$ are generically nonzero for the fixed aether theory.

The Einstein equation obtained from varying with respect to the metric takes the form
\begin{equation}\label{eqn:eelambda}
\eom_{ab}^{\ae} +\lambda u_a u_b = -\frac12 T_{ab},
\end{equation}
where $T_{ab}$ is the matter stress tensor, and  $\eom_{ab}^{\ae} = \dfrac{1}{\sqrt{-g}}\left.\dfrac{\delta S_{\ae}}{\delta g^{ab}}\right|_{u^c}$ is the metric variation of (\ref{eqn:aeaction}) holding $u^c$ fixed (see equation (\ref{eqn:aemetricvar})).  Equation (\ref{eqn:aethersource}) can be used to eliminate $\lambda$ in favor of $\mu$, in which case (\ref{eqn:eelambda}) becomes
\begin{equation} \label{eqn:frozenEE}
\eom^{\ae}_{ab}+\frac12 u^c\eom_c u_a u_b +\frac12 \mu u_a u_b= -\frac12T_{ab}.
\end{equation}
The left hand side contains the usual Einstein-aether metric equation terms 
plus an additional energy source coming from $\mu$.  
The aether internal energy is conserved, in the sense that $\tilde\mu u^a$ is a conserved current density: 
\beq
\nabla_a(\tilde\mu u^a)=\lie_u\tilde\mu = \lie_u(u^a\tilde\mu_a) = 0.
\eeq
Thus $\tilde\mu$ is like the density of a conserved mass that is carried along by the aether flow.  
The presence of this additional mass density that follows flows of $u^a$ motivates the 
name {\it ponderable aether} for these theories.

\subsection{Constraints}\label{sec:constraints}
Einstein-aether theory is a diffeomorphism invariant theory, which 
implies
that some of its equations of motion are constraints rather than dynamical equations.  These constraints place restrictions on the allowed initial values in the theory.

The constraints involve a combination of the metric \eqref{eqn:eelambda}
and aether \eqref{eqn:aethersource}
equations of motion \cite{Jacobson2011, Seifert2007a}.  
For
a given time coordinate $t$ 
(a scalar function on spacetime) the constrained quantities in the absence of matter are
\begin{equation}\label{eqn:constraints}
\tilde\cnstr_b^{(t)} = \nabla_a t\left(2{\tilde\eom}\indices{^{\ae}^a_b} + u^a\tilde\eom_b\right).
\end{equation}
This combination of the equations of motion involves one fewer $t$-derivative than the other equations, hence their role as constraints on initial values.  

Since these quantities
involve the aether field equation, they will not vanish in the nondynamical aether theory.    Instead, the source density $\tilde\mu_b$ characterizes the constraint violation,
\begin{equation}\label{Ct}
\tilde\cnstr_b^{(t)} = -(\lie_u t)\tilde\mu_b.
\end{equation}
The evolution of the 
constraint violation in time is determined by 
\eqref{liemutilde}, which together with \eqref{Ct} implies
\begin{equation}\label{lieCt}
\lie_u \tilde\cnstr_b^{(t)} = -(\lie_u\lie_u t)\tilde\mu_b.
\end{equation}
For generic $t$ this is not
a simple evolution equation for $\tilde\cnstr^{(t)}_b$; however, 
if $t$ is a proper time coordinate $\tau$ for $u^a$, i.e.\ $u^a\nabla_a t=1$, then the right hand side of \eqref{lieCt} vanishes. 
In an aether-adapted coordinate system, where the aether has components
\begin{equation}
u^\alpha = (1,0,0,0),
\end{equation}
the constraint quantities satisfy
\begin{equation}
\frac{\partial}{\partial \tau}\tilde\cnstr_\beta^{(\tau)} = 0;
\end{equation}
i.e.\ they are simply constant along the flow of $u^a$.

Another situation where the time evolution of the constraint quantities simplifies
is when $t=\eta$ is the parameter for a vector field $\zeta^a=Nu^a$ proportional to 
the aether, with $N$ a scalar function and  $\zeta^a\nabla_a \eta=1$. 
This example is relevant in  cosmological applications, in which $\eta$ is the conformal time
and $N=a$ is the scale factor.
In this case the constraint evolution is given by
\begin{align}
\lie_\zeta \tilde\cnstr_b^{(\eta)} & =-(\lie_\xi N^{-1})\tilde\eom_b -
N^{-1}\lie_\xi\tilde\eom_b \nonumber \\
&=-(N\lie_u N^{-1}+N^{-1}\lie_u N)\tilde\eom_b - \lie_u\tilde\eom_b -u^a\tilde\eom_a N^{-1}\nabla_b N \nonumber\\
&=- \tilde\mu\nabla_b\ln N. \label{eqn:logN}
\end{align}
Contracting this equation with $\zeta^a$ yields the evolution of the scalar constraint,
\beq\label{eqn:scalaretaconstr}
N\lie_\zeta(u^a\tilde\cnstr_a^{(\eta)}) = 0.
\eeq
In a coordinate system where $\zeta = \partial_\eta$ is the time evolution vector, this
equation reduces to an overall $\eta$-derivative,
\begin{equation} \label{eqn:hconstr}
\frac{\partial}{\partial\eta}\left(u^\alpha\tilde\cnstr_\alpha^{(\eta)}\right) = 0,
\end{equation}
and allows for immediate integration.  Furthermore, if $N$ depends only on $\eta$ and 
not on the spatial coordinates, the spatial components of (\ref{eqn:logN}) vanish.  This
means the spatial constraints also satisfy the simple evolution law,
\beq\label{eqn:momconstr}
\frac{\partial}{\partial \eta }\tilde\cnstr_i^{(\eta)} = 0.
\eeq

When we examine cosmological perturbations in section \ref{sec:scalarpert}, we choose
a gauge where the aether is aligned with the time evolution vector $\partial_\eta$ and 
receives no linear perturbations.  In this case, $N = a(\eta)$ is independent of the spatial
coordinates, and indeed we find a time derivative of a constraint in one of the linearized
field equations (\ref{eqn:CK}).  Also, the differentiated constraint (\ref{eqn:scalaretaconstr})
appears in equation (\ref{eqn:aetherscalarconstr}) in the case that no additional matter
other than the ponderable aether is present in the universe.  

\subsection{Relation to Lorentz-violating dark matter}\label{sec:lvdm}

Before proceeding with the phenomenology of these theories, it is worth considering  what physical situations could give rise to a frozen aether theory.  In \cite{Jacobson2014}, we showed that the frozen aether can be thought of either as a fundamental background structure on spacetime, or as a type of higher-derivative field theory for four scalar fields, $(\psi, \vphi^1, \vphi^2, \vphi^3)$.  In this section, we will describe a different situation, where the frozen aether appears as a limiting case of a dark matter theory coupled to dynamical Einstein-aether theory.  

The effects of coupling the aether to dark matter have been explored in \cite{Blas2012a}.  In that work, the ``pull-back formalism'' \cite{Andersson2007} (see also recent applications in \cite{Dubovsky2006,Dubovsky2012}) was employed in order to give the dark matter fluid a Lagrangian description.  The basic elements of this construction are three Euler potentials $\phi^I$, $I=1,2,3$, whose gradients $\phi^I_b\equiv\nabla_b\phi^I$ define the fluid's $4$-velocity $v^a$ via
\begin{align}
v^a &= \frac{J^a}{|J|}, \\
J^a & = \frac{1}{3! \sqrt{-g}}\tilde{\varepsilon}^{bcda}\phi^I_b\phi^J_c\phi^K_d\varepsilon_{IJK}.
\end{align}
Here, $|J|\equiv\left(J^a J^b g_{ab}\right)^{1/2}$, $\tilde{\varepsilon}^{bcda}$ is the anti-symmetric tensor density normalized such that $\tilde\varepsilon^{0123}=-1$, and $\varepsilon_{IJK}$ is a totally antisymmetric set of coefficients with $\varepsilon_{123}=1$.  

The action for a perfect fluid respects a symmetry under volume-preserving diffeomorphisms of the scalars, $\phi^I\mapsto\bar\phi^I(\phi^J)$, $\det\left(\dfrac{\partial \bar\phi^I}{\partial \phi^J}\right)=1.$  The simplest scalar invariant under this transformation is $|J|$, and the fluid Lagrangian is constructed to be a function of this scalar.  The form of this function determines the fluid's equation of state, and it is straightforward to see that
\begin{equation}
S_{\text{DM}} = -m\int d^4x\sqrt{-g} |J|
\end{equation}
gives a stress tensor for pressureless dust.  

The lowest order scalar that includes coupling to the aether while maintaining the $\phi^I$-diffeo symmetry is $\gamma\equiv g_{ab}u^a v^b$.  The generalization of $S_\dm$ to include Lorentz-violating effects simply multiplies the Lagrangian by a function of $\gamma$,
\begin{equation}
S_{\text{LVDM}} = -m\int d^4x\sqrt{-g}{F(\gamma)}{|J|}.
\end{equation}
$F$ is further subject to the restriction $F(1)=1$ to recover the ordinary dark matter limit when the dark matter has no velocity relative to the aether.  When $\gamma-1$ is small, $F$ can be organized in an expansion,
\begin{equation}\label{eqn:F}
F(\gamma)=  1+Y(\gamma-1)+\ldots
\end{equation}  
where $Y=F'(1)$ is a constant.  The limit we are interested in is a tight coupling between the aether and dark matter, enforced by sending $Y\rightarrow\infty$.  In this limit, $u^a$ and $v^a$ tend to align, causing $(\gamma-1)\rightarrow 0$, precluding the need to consider additional terms in the expasion (\ref{eqn:F}).  We will truncate $F$ at order $(\gamma - 1)$ in what follows.  

Parameterizing the misalignment of $v^a$ by
 \begin{equation}
 v^a = \gamma u^a + v_\perp^a,
 \end{equation}
 where $u\cdot v_\perp = 0$, we find that since $v^a$ is unit, $\gamma$ is related to $v_\perp$ by
 \begin{equation}
 \gamma = \sqrt{1+|v_\perp|^2}\approx 1 + \frac{|v_\perp |^2}{2}.
 \end{equation}
Thus for small $v_\perp^a$, $\gamma$ will deviate from $1$ as the square of the perpendicular velocity $|v_\perp|$.  

The aether equation of motion receives a correction due to the coupling to the dark matter,
\begin{equation}
\eom_a-2\lambda u_a -\mu_{\dm} Y v_a = 0,
\end{equation}
where we have defined $\mu_\dm =  m |J|$.  The $u^a$ component determines $\lambda$, 
\begin{equation}\label{eqn:lvdmlambda}
\lambda = \frac12u^a\eom_a -\frac12\mu_\dm Y\gamma,
\end{equation}
and the $v_\perp^a$ component gives the relation 
\begin{equation}
\frac{1}{|v_\perp|}v_\perp^a\eom_a = -\mu_\dm Y \gamma |v_\perp|.  
\end{equation}
Since the LHS of this equation should generically be the same order as $\mu_\dm$, this relation shows that $|v_\perp|\sim \mathcal{O}(1/Y)$.  This also implies that $\gamma=1+\mathcal{O}(1/Y^2)$. 

Finally we examine the metric variation:
\begin{equation}
\eom_{ab}^{\ae} +\lambda u_a u_b +\frac{\mu_\dm}{2}\left( (1-Y) v_a v_b +2Yu_{(a}v_{b)}  \right)
\end{equation}
Using (\ref{eqn:lvdmlambda}) to eliminate $\lambda$ gives
\begin{equation}
\eom^{\ae}_{ab}+\frac12 u^c\eom_c u_a u_b +\Delta \eom_{ab},
\end{equation}
where $\Delta\eom_{ab}$ describes the deviation from Einstein-aether theory.  The expression for this tensor is 
\begin{align}
\Delta&\eom_{ab} = \frac{\mu_\dm}{2}\Big(\left(\gamma^2(1-Y)+\gamma Y\right) u_a u_b \nonumber \\
&+\big(Y(1-\gamma)+\gamma\big) 2 u_{(a}{v_\perp}_{b)}   +\left(1-Y\right){v_\perp}_a{v_\perp}_b\Big)
\end{align}

Now taking the limit $Y\rightarrow\infty$, and using the derived $Y$-dependence of $\gamma$ and $v_\perp^a$, we find in the limit the only finite piece is
\begin{equation}
\Delta \eom_{ab} \rightarrow \frac12 \mu_\dm u_a u_b,
\end{equation}
with all other terms falling off as $\mathcal{O}(1/Y)$ or faster.  We thus recover the frozen aether Einstein equation (\ref{eqn:frozenEE}) with the dark matter energy density $\mu_\dm$ playing the role of $\mu$.  

This limit is useful because it allows results about Lorentz-violating dark matter from \cite{Blas2012a} to be applied to the frozen aether theory.  In particular, the analytical solutions for cosmological perturbations obtained for Lorentz-violating dark matter match on to the solutions we consider in this work when the $Y\rightarrow\infty$ limit is taken.

\section{Small-scale phenomenology}\label{sec:smallpheno}

Here we investigate some astrophysical effects of the frozen aether by considering waves in Minkowski space, the Newtonian limit, and spherical symmetry.  The most notable effect comes from ponderable aethers (those with $\mu\neq 0$) in the Newtonian limit, where $\mu$ plays the role of a dark matter source.  On the other hand, we find that nonzero $\mu$ precludes static solutions.  The nonexistence of static, spherically symmetric solutions for ponderable aethers is a significant challenge, and leads to a puzzle about the final state of gravitational collapse.  

\subsection{Linearized waves} \label{sec:linearwaves}
Linearized perturbations around Minkowski space closely resemble the results in Einstein-aether theory \cite{Jacobson2004,Foster2006}.  There are five massless wave solutions, corresponding to spin-2, spin-1, and spin-0 excitations.  In the background, the aether is aligned with a Minkowski time translation killing vector, and $\mu_a=0$.  Hence when perturbations are considered, the integration constants $\mu_a$ must be of the same perturbative order as the metric and aether perturbations.

Using the SVT decomposition of metric and aether perturbations
described in Appendix \ref{sec:pertvars}, we find that the scalar variable 
$\gis$ obeys the sourced wave equation
\begin{equation}\label{eqn:spin0}
\square_0\gis = 4\pi G_N\left( \rho +\mu+3 c_a p-3c_a \frac{2+c_\theta}{2c_\sigma+c_\theta}\triangle \pi_L \right).
\end{equation}
Here, $\square_0\equiv s_0^{-2}\partial_t^2-\triangle$ is the spin zero wave operator with wave speed 
\begin{equation}\label{eqn:s02}
s_0^2  = \frac{(c_\theta+2c_\sigma)(2-c_a)}{3c_a(1-c_\sigma)(2+c_\theta)},
\end{equation}
 $\triangle\equiv \delta^{ij}\partial_i \partial_j$ is the spatial Laplacian,
\begin{equation}\label{eqn:GN}
G_N \equiv \dfrac{G_0}{1-\frac{c_a}{2}}
\end{equation}
 is the renormalized Newton constant for the Einstein-aether Newtonian limit \cite{Carroll2004}, $\rho$ is the matter energy density, and the pressure $p$ and anisotropic stress $\pi_L$ are defined by decomposing the scalar parts of the spatial matter stress  tensor by,
\begin{equation}
T^{(m)}_{ij} = p\delta_{ij} + \left(\partial_i\partial_j-\frac13\delta_{ij}\triangle\right) \pi_L.
\end{equation}

Equation (\ref{eqn:spin0}) matches the sourced spin-$0$ wave equation from reference \cite{Foster2006} except for the additional term coming from $\mu$.  The evolution of $\mu$ given by $\lie_u\tilde\mu = 0$ becomes at linear order in perturbations,
\begin{equation}
\partial_t\mu=0.
\end{equation}
Thus, $\mu$ is a strange type of source that is constant in time.  The general solution for these wave modes will be the same as in Einstein-aether theory, except the equilibrium value of 
each Fourier mode of $\gis$ will be shifted by a $\mu$-dependent constant.  This zero frequency shift would be largely undetectable away from the source.

The spin-$1$ analysis produces a similar result.  The vector variable $\giv_i$ defined in (\ref{eqn:vgiv}) satisfies a wave equation,
\begin{equation}
\square_1\giv_i=\frac{16\pi G(c_\sigma S_i +\kappa_i)}{1-(1-c_\sigma)(1-c_\omega)},
\end{equation}
where $\square_1 = s_1^{-2}\partial_t^2-\triangle$, the spin-$1$ wave speed is 
\begin{equation}
s_1^2 = \frac{1-(1-c_\sigma)(1-c_\omega)}{2c_a(1-c_\sigma)},
\end{equation}
$S_i$ is the vector part of $T^{(m)}_{0i}$, and $\kappa_i$ is the vector part of the source  $\mu_i$.  This again agrees with reference \cite{Foster2006} (see erratum), modulo the extra integration constant $\kappa_i$.  As with the scalar integration constant, $\kappa_i$ is constant in time, so again only produces a shift in the zero point value of $V_i$. 

The spin-$2$ equations of motion are exactly the same as in Einstein-aether theory, since there is no tensor component to $\mu_a$.

\subsection{Newtonian limit}\label{sec:newtonian}

The Newtonian limit is recovered from the linearized equations by requiring slow motion of sources so that time derivatives in the linearized equations are suppressed by powers of $v\ll 1$.  There are two scalar gravitational potentials, given by $\gis$ and $\gih$ from (\ref{eqns:minkpert}).  Equation (\ref{eqn:spin0}) then reduces to a Poisson equation
\begin{equation} \label{eqn:poisson}
\triangle \gis = -4\pi G_N (\rho+\mu),
\end{equation}
with $G_N$ defined in (\ref{eqn:GN}).  The $0$ component of the constraint (\ref{eqn:constraints}) gives a relation for the other gravitational potential,
\begin{equation}
\triangle(-2\gis-c_a\gih) = 8\pi G (\rho + \mu)
\end{equation}
which then implies that $\gih$ satisfies the same equation (\ref{eqn:poisson}) as $\gis$, so
\begin{equation}
\gis=\gih.
\end{equation}

The Newtonian gauge choice $B=U=0$ shows that $\gis$ and $\gih$ function as the Newtonian potentials (the remaining scalar potential $\giv$ which functions as a velocity potential for the aether is only determined at $0.5$ post-Newtonian order).  Equation (\ref{eqn:poisson}) demonstrates that $\mu$ will contribute an extra matter component in the Newtonian limit, which could be interpreted as the dark matter.  In general the distribution of 
$\mu$ need not bear any simple relation to the density of ordinary matter $\rho$.  The value of $\mu$ will depend on the details of how gravitational collapse occurred from the cosmological evolution. 
This provides a possible motivation for exploring whether $\mu$ can be an astrophysically viable explanation for dark matter.

\subsection{Static spherical symmetry}\label{sec:sphere}

Static solutions for ponderable aethers are somewhat problematic, due a tendency for the aether to align with the timelike Killing vector $\xi^a$.  If it does not align, there will generically be a flux of $\mu$ energy density across timelike hypersurfaces generated by flows of $\xi^a$.    This will lead to accumulation of $\mu$ inside of closed spatial regions, which will presumably  lead to  a time-varying source in the Einstein equations.  Hence, nonaligned ponderable aethers are inconsistent with the existence of a static Killing vector.  

However, when the aether is aligned with $\xi^a$, $\tilde\mu$ is required to 
vanish by symmetry.  To see this, we write $u^a = N^{-1}\xi^a$, and  from 
(\ref{eqn:lieaethereom}) and (\ref{eqn:aethersource}) find
\begin{equation}\label{eqn:stationarylie}
0=\lie_u\tilde\mu_a=  N^{-1}\lie_\xi \tilde\mu_a + \tilde\mu_a\lie_\xi N^{-1} -\tilde\mu \nabla_a \ln N.
\end{equation}
Since $\xi^a$ is a Killing vector, $\lie_\xi\tilde\mu_a=0$ and $\lie_\xi N^{-1}=0$.  Also, for $u^a$ aligned with $\xi^a$, the acceleration is given by $a_a = -\nabla_a \ln N$.  Since this will not vanish except in the case of Minkowski space, (\ref{eqn:stationarylie}) implies $\tilde\mu=0$. 

Imposing the additional restriction of staticity is enough to set the perpendicular source density $\tilde\mu^\perp_a$ to zero as well.  The alignment of  $u^a$ with a static Killing vector implies $\theta=0$, $\sigma_{ab}=0$, and  $\omega_{ab}=0$.  Also, the relation $\lie_u a^c = - u^c a^2$ holds in this case, so examining the form of $\eom_b$ given by equation (\ref{eqn:aethereom}), we find
\begin{equation}
\eom_b \propto a^2 u_b,
\end{equation}
so in particular, 
\begin{equation}
\tilde\mu_b^\perp = \tilde\eom_b^\perp = 0.
\end{equation}
This eliminates the possibility of new static, spherically symmetric solutions, outside of those that already exist in Einstein-aether theory where $\tilde\mu_a = 0$ \cite{Eling2006b}.  Note that a similar effect has been found in projectable \Horava gravity, where accretion of the dark matter integration constant prevents static stellar solutions \cite{Izumi2010}.

The only possibility left is to drop the assumption that $u^a$ aligns with $\xi^a$.  The general argument at the beginning of this section would still apply, meaning that these solutions would need $\tilde\mu=0$.   However, we no longer can conclude that $\tilde\mu_b^\perp$ kinematically vanishes, so it may be possible to find fixed threading solutions that generalize the Einstein-aether spherically symmetric solutions \cite{Eling2006b, Eling2006a}.  This prospect is left to future work. 

\section{Cosmological phenomenology}\label{sec:cosmopheno}

In this section we look at the cosmological implications of the sources $\tilde\mu_a$.  The most interesting features come from ponderable aethers, since the background value of $\tilde\mu_a^\perp$ will vanish by isotropy.   The scalar source  $\tilde\mu$ functions as a dark matter component, affecting the background FLRW solutions.
 This is an interesting result, and  motivates investigating whether it can reproduce other dark matter phenomenology.  

The first issue that must be addressed is determining what fraction of the matter content today could be attributed to  $\mu$.   This requires determining initial conditions for $\mu$, about which, unfortunately, our theory does not have much to say.  It seems necessary
to simply take the initial value of $\mu$ as a set of parameters
defining the theory.  Given a choice of initial condition, we can estimate the size of $\mu$ 
today by noting that $\mu\propto a^{-3}$.  Then the fraction of matter comprised of $\mu$ today is related to the initial density $\mu_\text{inf}$ by 
\begin{equation}
\Omega_{\mu0} = \mu_{\text{inf}} \,e^{-3N} \left(\frac{a_0}{a_\text{reh}}\right)^{-3}\frac{1}{\rho_\text{crit}}.
\end{equation}
For an inflationary Hubble rate of $H\sim 2\times10^{16} \,\text{GeV}$, a temperature of reheating to be  $T_\text{reh} \sim2\times 10^{17}\,\text{GeV}$, and an initial value for the source to be the same order as the vacuum energy density during inflation $\mu_\text{inf}\sim\rho_\Lambda\sim 2\times10^{70}\,(\text{GeV})^4$,  we estimate a  value today of $\Omega_{\mu0}\sim 10^{-53}$.  This negligibly small value for $\Omega_{\mu0}$ effectively reduces the theory to dynamical Einstein-aether theory.  

Since the primordial value $\mu$ would be washed out by inflation, the only hope for there to be observational consequences is for $\mu$ to somehow be generated at the end of inflation, after which it subsequently satisfies the classical conservation law $\lie_u\tilde\mu = 0$.  However, this situation is fairly implausible as well, since we would need $\Omega_{\mu,\text{reh}}\sim10^{-26}$  to prevent it from dominating before matter-radiation equality.\footnote{However, see \cite{Chamseddine2013, Mirzagholi2014} for proposals for generating
$\mu$ at matter-radiation equality in the mimetic theory, and \cite{Mukohyama2009} for
the generation of dark matter in projectable \Horava gravity.}

Putting aside the issue of generating the appropriate value of $\mu$ primordially, there is yet another issue with using $\mu$ as a dark matter candidate: it does not lead to structure formation.  Perturbations to FLRW spacetimes dominated by $\mu$ will contain two vector modes and one scalar mode, just like the waves in flat space form section \ref{sec:linearwaves}.  If $\mu$ is the dominant form of matter in the universe, it will be solely responsible for the formation of structure through gravitational collapse.  
However, it is already clear that this will be problematic.  Ordinary cold dark matter, as a pressureless fluid, has growing modes at all scales during matter domination.  This is not the case for the ponderable aether.  Although its stress tensor behaves like a pressureless fluid in the background spacetime, in general $\delta p_\aeth\sim s_0^2 \delta\rho_\aeth\neq0$, where $s_0^2$, defined in (\ref{eqn:s02}), is the spin-$0$ mode speed for Einstein-aether waves. Hence, there exists a characteristic Jeans length $\lambda_J^{\ae}\sim s_0/H$ (with 
Hubble rate $H$),  below which the aether pressure will prevent gravitational collapse.  Since phenomenological constraints coming from the absence of gravitational \v{C}erenkov radiation require $s_0\geq 1$ \cite{Elliott2005}, this means structure would not form on any scales smaller than the Hubble length.  

Thus, this model still requires an ordinary dark matter fluid in order for scalar perturbations to grow. 
The main effect of $\mu$ in the presence of ordinary matter is that it changes the relationship between $\mathcal{H}$ and the background matter energy density $\rho$.  Thus when $\Omega_\mu$ is small, the cosmological perturbations will be identical to dynamical Einstein-aether theory, which have been studied previously \cite{Lim2005, Li2008,Armendariz-Picon2010, Sierra2011}.   Ponderable aethers deviate from Einstein-aether theory only  when the universe enters a matter-dominated phase, where $\Omega_\mu$ may be a significant fraction of unity.  We will therefore focus on this phase of the universe.    The analysis of scalar perturbations in this situation reveals growing mode on superhorizon scales, with a growth rate depending on the value of $\Omega_\mu$.  This allows us to phenomenologically constrain $\Omega_\mu$ to be less than $\mathcal{O}(10^{-2})$, precluding its interpretation as a significant source of  dark matter.

\subsection{FLRW solutions}\label{sec:frw}
In ordinary Einstein-aether theory, the only effect of the aether on flat FLRW solutions is to rescale the gravitational constant appearing in the Friedmann equations to $G_\text{cos  } = G_0/(1+\frac{c_\theta}{2})$ \cite{Carroll2004}.  The ponderable aether gives rise to a new effect; the source  $\mu$ contributes additional energy density to the Friedmann equation that behaves like pressureless dust, with $\mu\propto a^{-3}$.  

Working with a flat FLRW metric in conformal time coordinates (\ref{eqn:flrwmetric}), isotropy requires that $u^a$ be  aligned with the time translation vector $\partial_\eta$, with components
\begin{equation}
u^\alpha = a^{-1}\delta^\alpha_\eta.
\end{equation}
This form is especially convenient for computing Lie derivatives $\lie_u$ and other objects related to the aether.  
For example, the expansion $\theta$ is 
\begin{equation}
\theta = \frac{\lie_u\sqrt{-g} }{\sqrt{-g}}= a^{-4} (a^{-1}\partial_\eta a^4+a^4\partial_\eta a^{-1}) = 3 \frac{\hc}{a},
\end{equation}
with $\hc\equiv a'/a$, and prime denotes differentiation with respect to $\eta$.  

Isotropy imposes that $\omega_{ab}=\sigma_{ab}=0$, $a_a = 0$, so that the expansion $\theta$ entirely characterizes the aether.    The aether variation (\ref{eqn:aethereom}) takes the simple form
\begin{equation}
\eom_a = \frac{1}{8\pi G_0}\frac{c_\theta}{3} \nabla_b\theta,
\end{equation}
the metric variation (\ref{eqn:aemetricvar}) depends only the expansion piece $\mathcal{X}_{ab}$ defined in equation (\ref{eqn:Xab}), $\eom_{ab}^{\ae} = \frac{-1}{16\pi G_0}(G_{ab}+\frac{c_\theta}{3}\mathcal{X}_{ab} )$.  The fixed aether Einstein equation (\ref{eqn:frozenEE}) now reads
\begin{align}
 \frac{\left(1+\frac{c_\theta}{2}\right)}{a^2}\left[2(\hc^2-\hc')u_au_b+(2\hc'+\hc^2)g_{ab}\right]  
= 8\pi G_0( T_{ab}+\mu u_a u_b).
\end{align}
This gives the two Friedmann equations governing the dynamics,
\begin{subequations} \label{eqns:friedmann}
\begin{align}
a^{-2}\hc^2 &= \frac{8\pi G_{\text{cos}}}{3} (\rho+\mu), 
\label{eqn:friedmann} \\
a^{-2} \hc' &= -\frac{4\pi G_\text{cos}}{3}(\rho+3p),
\end{align}
\end{subequations}
where $\rho$ and $p$ are the matter energy density and pressure.

Hence, the novel feature of this theory compared to Einstein aether theory is the extra energy component appearing in the Friedmann equation (\ref{eqn:friedmann}) coming from $\mu$.  Also the requirement that $\lie_u\tilde\mu = 0$ means the $\mu\propto a^{-3}$, just like 
pressureless dust.

In our analysis of perturbations around these backgrounds, we will focus on the matter dominated phase where $\mu$ can represent an appreciable fraction of the energy density of the universe.  We will use a simple model consisting of the ponderable aether and an additional dark matter fluid.  Since our focus will be on scalar perturbations, it is useful to assume that the dark matter fluid is irrotational.  To provide a Lagrangian description of such a fluid, we employ the ``angel dust''  model \cite{Lim2010}, consisting of a scalar field $\vphi$ and a Lagrange multiplier field $\rho$ with the action
\begin{equation}\label{eqn:matteraction}
S_\text{m} = \int d^4x \sqrt{-g}\,\frac{\rho}{2}\left( g^{ab}\nabla_a\vphi \nabla_b\vphi - 1\right).
\end{equation}
The $\rho$ equation of motion enforces that $v_a \equiv \nabla_a\vphi$ be unit, which also implies that $v^a$ is geodesic.  The stress tensor is that of pressureless dust,
\begin{equation}
T^\text{m}_{ab} = \rho v_a v_b.
\end{equation}

In homogeneous, isotropic cosmology, $u^a=v^a$, and since $p$ for the dark matter vanishes, the solution to the Friedmann equations (\ref{eqns:friedmann}) is a matter-dominated universe with $a\propto \eta^2$, $\mathcal{H} = 2/\eta$.  The dark matter fields satisfy 
\begin{subequations}\label{eqns:bgscalars}
\begin{align}
\vphi' &= a, \\
\rho &= \frac{3(1+\frac{c_\theta}{2})}{8\pi G_0}\frac{\mathcal{H}^2}{a^2} (1-\Omega_\mu) \propto a^{-3},
\end{align}
\end{subequations}
with $\Omega_\mu\equiv \frac{8\pi G_\text{cos} a^2}{3\hc^2} \mu$.  Note that during matter domination, $\Omega_\mu$ remains constant.

\subsection{Scalar perturbations}\label{sec:scalarpert}

Now we turn to deriving equations for linearized perturbations about this solution.  The metric and aether perturbations are parameterized using the SVT decomposition described in Appendix \ref{sec:pertvars}.  The equations of motion can be derived by expanding the actions (\ref{eqn:aetheraction}) and (\ref{eqn:matteraction}) about the background solution to quadratic order in the perturbative variables.  Varying the resulting quadratic action yields the linearized equations of motion.  

For the ponderable aether, we must also ensure that the linearized aether equation of motion is not imposed by this procedure.  These equations result from varying the aether perturbations $T$, $U$ and $U^i$.  To avoid enforcing these equations of motion, we can gauge-fix these quantities at the level of the action, so that their variations are not included.  Choosing $T=U=U^i=0$ simply selects the gauge where $u^\alpha = a^{-1}\delta^\alpha_0$.   This also allows us to solve the unit constraint for the $g_{00}$ metric perturbation, and gives $A=0$.  When this is substituted back into the action, the $\lambda$ term (\ref{eqn:lagrange}) drops out.

These conditions in fact do not fully fix the gauge; there are still coordinate transformations that leave this form of $u^\alpha$ invariant.  These correspond to shifts of the time coordinate along each of the aether threads, $t\mapsto t+f(x^i)$, and thread-preserving reparameterizations of the spatial coordinates, $x^i\mapsto \bar{x}^i(x^j)$.  Rather than trying to fully eliminate this residual gauge ambiguity, we can instead focus on gauge invariant combinations of the remaining nonzero variables.  These combinations are given in equations (\ref{eqns:gi}), and in our choice of gauge take the form
\begin{subequations}
\begin{align}
\gis & \eqcoord -C-\hc B, \\
\gih &\eqcoord B' +\hc B,\\
\giv &\eqcoord E', \\
\gix &\eqcoord -\hc\left(\left(\frac{C}{\hc}\right)'+C\right),
\end{align}
\end{subequations}
and $\eqcoord$ denotes equality only in this gauge.  

The matter field $\vphi$ and $\rho$ also receive perturbations,
\begin{subequations}
\begin{align}
\rho &= \bar\rho + \delta\rho  \\
\vphi&=\bar\vphi + \delta\vphi,
\end{align}
\end{subequations}
where the bar denotes the background value of the fields, given by (\ref{eqns:bgscalars}).  
The useful gauge-invariant quantities associated with these are
\begin{subequations}
\begin{align}
\gim&\equiv\hc\left(B+U+\frac{\delta\vphi}{\bar\vphi '}\right)\eqcoord\hc\left(B+\frac{\delta\vphi}{\bar\vphi '}\right),  \\
\giz&\equiv \gis + \gim = -C+\hc\frac{\delta\vphi}{\bar\vphi'}, \\
\gid &\equiv \frac{\delta\rho}{\bar\rho}+3C.
\end{align}
\end{subequations}
The variable $\giz$ has the geometric interpretation of the curvature perturbation on surfaces of constant $\vphi$.  When we consider only scalar perturbations, the aether is orthogonal to surfaces of constant $B+U$, so that $\gis$ gives the curvature perturbation on these surfaces.  The variable $\gim$ is then related to the difference in conformal time between these surfaces \cite{Armendariz-Picon2010}.  The final variable $\gid$ is simply the matter density contrast in the flat-slicing gauge, where $C=0$.

It is convenient to work with the spatial Fourier transform of the perturbative variables, defined by
\begin{equation}
X(\mathbf{x}, \eta) = \int\frac{d^3 k}{(2\pi)^{3/2}} e^{i \mathbf{k}\cdot\mathbf{x}} X_{\mathbf{k}}(\eta).
\end{equation}
From now on, we will work with the Fourier-transformed variables $X_{\mathbf{k}}$ and drop their explicit dependence on $\mathbf{k}$. 

The aether action (\ref{eqn:aetheraction}) is now expanded to quadratic order.  The details of this calculation for the aether quantities are given in Appendix \ref{sec:pertcalcs}.  Using the background equation $2\hc '+\hc^2=0$ and integrating by parts,  the action takes the form
\begin{align}
S^{\ae}_{(2)} &= \frac{-1}{16\pi G_0}  \int d\eta\, a^2 \left[ \frac{(2+c_\theta)}{12} {h'}^2 - \frac{2(1-c_\sigma)}{3}(k^2 E')^2 
 -c_ak^2a^{-2}{(aB)'}^2  -2k^2(\gis^2-2\gis\gih )  \right], \label{eqn:quadaether}
\end{align}
where 
\beq
h=6C-2 k^2E.
\eeq
The matter action (\ref{eqn:matteraction}) at quadratic order is 
\begin{align}
S^\text{m}_{(2)}  =\frac{3}{16\pi G_0}& \int  d\eta\, a\hc^2(1-m)\left[ \frac{\delta\rho}{\bar\rho} \delta\vphi'  - k^2\frac{\bar\vphi'}{2}\frac{M^2}{\hc^2}    
+\frac12 h\delta\vphi'  +\frac12\bar\vphi'\left( \frac{(\delta\vphi')^2}{(\bar\vphi')^2} + k^2 \frac{\delta\vphi^2}{(\bar\vphi')^2}\right) \right]  ,  \label{eqn:quadmatter}
\end{align}
with 
\begin{equation}
m\equiv-\frac{c_\theta}{2}+\Omega_\mu\left(1+\frac{c_\theta}{2}\right).
\end{equation}
   Here we have used equation (\ref{eqns:bgscalars}) to express $\bar\rho$ and $\bar\vphi'$ in terms of $a$ and $\hc$.  

The total quadratic scalar action $S^{\ae}_{(2)}+S^\text{m}_{(2)}$ is thus a function of five independent variables $(h,E,B,\delta\rho, \delta\vphi)$, since $\gis$, $\gih$, and $\gim$ are simply functions of these.  We expect there to be two dynamical modes for this system corresponding to the aether scalar mode and the angel dust mode.  The equation from varying $\rho$ takes the form of a constraint, and from the discussion in section \ref{sec:constraints} there are two other constraints with undetermined integration constants, given by equation (\ref{eqn:hconstr}) and the scalar part of equation (\ref{eqn:momconstr}).  These will reduce the original five variables down to the two dynamical ones.  

The $\delta\rho$ variation gives the first constraint,\footnote{The gauge-invariant
version of this equation is $\frac{\delta\vphi'}{\bar\vphi'}-A=0$, but $A=0$ for our
gauge choice.} 
\begin{equation}\label{eqn:phiconst}
\delta\vphi' = 0.
\end{equation}
This helps simplify the equations; for example, the $h\delta\vphi'$ term in (\ref{eqn:quadmatter}) no longer contributes to the $h$ equation.  Varying now with respect to $\delta\vphi$ yields another constraint, which  in terms of gauge invariants  reads
\begin{equation} \label{eqn:deltaF}
k^2 \left(\giv-\frac{\gim}{\hc}\right) -\gid' =0.
\end{equation}
Since $\gid$ does not enter any of the other equations, this equation simply determines it 
once the solutions for $M$ and $V$ are known.

The remaining parts of the action now only depend on three variables $h$, $E$ and $B$.  It is more convenient to change variables from $h$ and $E$ to
\begin{align}
C &= \frac13 k^2 E+\frac16 h, \\
K &= \frac13 k^2 E - \frac16h.
\end{align}
After doing this, the  $K$-dependent terms of the action take the form (from now on we ignore the overall $1/(16\pi G_0)$ prefactor, which is common to all terms) 
\begin{align}
S^K = \int d\eta\, a^2\frac34\left((c_\theta+2c_\sigma){K'}^2-2(4+c_\theta-2c_\sigma)C'K'\right),
\end{align}
Here we see that $K$ is cyclic, so its equation of motion will be a total time derivative.  This is in fact 
the constraint expected from equation (\ref{eqn:momconstr}), 
which after integrating contains an undetermined constant $\mu_s$ corresponding to a source density,
\begin{equation}\label{eqn:CK}
(c_\theta+2c_\sigma) K' - (4+c_\theta-2c_\sigma)C' = a^{-2}\mu_s.
\end{equation}
This can be cast into a gauge-invariant form, using $K' = \frac23 k^2 \giv - C'$, and 
$C' = -\gix -\frac32\hc C$, which gives
\beq\label{eqn:gidiffconstr}
\frac23(c_\theta+2c_\sigma)k^2 \giv + 2(2+c_\theta)\gix=  \left(a^{-2}\mu_s-3
(2+c_\theta)\hc C\right).
\eeq
The gauge-invariance of the right hand side follows from viewing $\mu_s$ as arising from 
the scalar part of the 
spatial component of the constraint violation, 
${\tilde\cnstr^{(\eta)}_{i}} = \partial_i\mu_s$.
Under a gauge transformation generated by $\xi^\alpha$, this changes by 
$\delta\tilde\cnstr^{(\eta)}_i = 
\tilde\cnstr_0^{(\eta)}\partial_i\xi^0$, where the background value of $\tilde\cnstr_0^{(\eta)}$
is determined via the background Einstein equation to be $3(2+c_\theta)a^2\hc^2$. 
Gauge invariance then follows by noting that $C$ transforms as $\delta C = \hc \xi^0$. 
Equation (\ref{eqn:gidiffconstr}) determines $\giv$ once a solution $\giz$ is known,
since $\gix = -\frac{\hc}{a}\left(\frac{a\giz}{\hc}\right)'$, and $C$ is determined by $\giz$
up to a term proportional to $\delta\varphi$, which by (\ref{eqn:phiconst}) is just an
integration constant.

Equation (\ref{eqn:CK}) can be used to eliminate $K'$ from the  action (\ref{eqn:quadaether}).  When this is done, the $\mu_s$ piece decouples, and up to  total derivatives the action for the two dynamical modes, including the piece from (\ref{eqn:quadmatter}), can be written in terms of gauge invariants,
\begin{align}
S^{\text{dyn}}_{(2)} & = \int d\eta\left[\beta\hc^2{\left(\frac{a\giz}{\hc}\right)' }^2 + c_ak^2{\left(\frac{a\gim}{\hc}\right)'}^2 
+a^2k^2\left(2\giz^2+4\frac{\gim}{\hc} \giz' +3mM^2 \right) \vphantom{{\left(\frac{aZ}{\hc}\right)' }^2}\right],
\end{align}
where 
\begin{align}
\beta&\equiv\frac{6(2+c_\theta)(1-c_\sigma)}{c_\theta+2c_\sigma}.
\end{align}
From this action we can derive the two dynamical equations for $\giz$ and $\gim$.  Using the background equation $a=\eta^2$, they take the form
\begin{subequations}\label{eqns:scalarmodes}
\begin{align}
\beta\ddot{Z}+\frac{4\beta}{y}\dot{Z}+y\dot{M}-2Z+5M&=0,  \label{eqn:Z} \\
\frac{c_a}{2}\ddot{M}+\frac{3c_a}{y} \dot{M} -\frac2y \dot{Z}+\frac{3(c_a-2\mm)}{y^2} M &= 0, \label{eqn:M}
\end{align}
\end{subequations}
where $y\equiv k\eta$ and dots represent differentiation with respect to $y$. 

These equations describe how the two scalar modes evolve, and the dependence on $\Omega_\mu$ appears through the parameter $\mm$ in (\ref{eqn:M}).  An important set of solutions to these equations comes from the long wavelength limit $y\ll 1$, or equivalently $k_\text{phys} = k/a \gg H = \hc/a$, where equation (\ref{eqn:Z}) reduces to 
\begin{equation}\label{eqn:Zred}
\beta\left(\ddot{Z}+\frac4y \dot{Z}\right) = 0,
\end{equation}
and equation (\ref{eqn:M}) remains the same.  These equations now admit four independent power law solutions.  The first two solve (\ref{eqn:Zred}) with 
\begin{equation}
Z\propto y^{q_\pm}, \quad q_\pm = 0, -3.
\end{equation}
For the $q_+=0$ solution, $M$ is simply given by $M=0$.  For  $q_-=-3$, $M$ obeys the same power law as $Z$, with an amplitude determined by equation (\ref{eqn:M}) to be
\begin{equation}
\frac{M_-}{Z_-} = \frac{1}{m}.
\end{equation}
Then using equations (\ref{eqn:CK}) and (\ref{eqn:deltaF}), we find that the  matter density contrast $\delta^F$ evolves as $\delta^F\propto y^{q_\pm + 2}$, so that the growing mode satisfies $\delta^F\propto a$.  This growth rate is the same as cold dark matter during matter-domination.  Thus, we see  that the matter decouples from the scalar aether perturbations on superhorizon scales.  Once the mode enters the horizon, it will generically mix with the aether perturbations and lead to decaying oscillatory solutions.  

The other two solutions set $Z=\dot{Z}=0$, which is preserved by (\ref{eqn:Zred}).  Then (\ref{eqn:M}) has power law solutions $M\propto y^{p_\pm}$, with 
\begin{subequations} \label{eqns:exponent}
\begin{align}
p_\pm &= \frac52\left(-1\pm\sqrt{1+\frac{24}{25}\frac{\Omega_\mu(2+c_\theta)-c_\theta-c_a}{c_a}  }\, \right)  \\
&=  \frac52\left(-1\pm\sqrt{1+\frac{48 G_0}{25 c_a} \left(\frac{1}{G_N}-\frac{1-\Omega_\mu}{G_{\text{cos}}  } \right) } \, \right) \label{eqn:expG}
\end{align}
\end{subequations}
Results that reduce to this exponent when $\Omega_\mu=0$ have previously been derived in Einstein-aether theory \cite{Armendariz-Picon2010}, generalized Einstein-aether theory \cite{Zlosnik2008}, and \Horava gravity \cite{Kobayashi2010a}.  For nonzero values of $\Omega_\mu$, these exponents can also be derived from the analysis of Lorentz-violating dark matter \cite{Blas2012a},  by taking the parameter $Y\rightarrow\infty$, as described in section \ref{sec:lvdm}.\footnote{
Note that the exponents in (\ref{eqns:exponent}) diverge in the limit that $c_a\rightarrow0$,
and, as explained in section \ref{sec:isw},  small values of $c_a$ are phenomenologically
preferred.   On the other hand, if we set $c_a=0$
in the action, the $\gim$ field equation (\ref{eqn:M}) becomes a constraint relating $\gim$
to $\giz$, eliminating one of the scalar modes.  Hence, we see the $c_a\rightarrow0$
limit of (\ref{eqns:exponent}) does not match 
the $c_a=0$ solutions.  This is related to the strong-coupling issues
that arise in \Horava gravity and Einstein-aether theory in the limit that the parameters
$c_i$ become small \cite{Papazoglou2010, Blas2010a, Withers2009}.  In particular, 
the effective field theory becomes strongly coupled at energy scales of the order 
$\Lambda =\sqrt{c_a} M_P$.  
For \Horava gravity, the higher spatial derivative terms can provide
a UV completion of the IR theory defined by the action
(\ref{eqn:aetheraction}) \cite{Blas2010a}.  It is unknown whether a UV completion exists
for Einstein-aether theory.  Nevertheless, the strong coupling scale $\Lambda$
can remain close
to the Planck scale even for relatively small values of $c_a$.
}

Another interesting set of solutions occurs when the universe is dominated by the ponderable aether, i.e.\ $\Omega_\mu\rightarrow1$,  since in this case the order of the equations reduces to a single dynamical scalar.  To see this, we add (\ref{eqn:Z}) to $\frac12 y^2$ times (\ref{eqn:M}) and multiply by $y$,  resulting in
\begin{align}
\frac{d}{dy}\left[ \frac{\beta}{y^2}\frac{d}{dy}(y^3\giz) -y^2\giz +\frac{c_a}{4}\frac{d}{dy}(y^3 \gim) +y^2\gim \right] 
- 3(\mm-1)yM=0, \label{eqn:aetherscalarconstr}
\end{align}
which can be immediately integrated when $\mm=1$.  This gives a constraint equation that can be written entirely in terms of $\gis$ and $\gih$, and the integration constant resulting from this procedure is simply the perturbation $\delta\mu$ to the source  $\mu$.  

Solving the constraint for $\gih$, and inserting the result into (\ref{eqn:Z}) results in a single evolution equation written entirely in therms of $\gis$,
\begin{align}
y^2\ddot{\gis} +\left[4+\frac{2\xg}{y^2+\xg}\right] y\dot{\gis} + \left[s_0^2y^2-\frac{20}{c_a}+\frac{8(3\xs+\xg)}{c_a(y^2+\xg)}  \right] \gis 
=\left[ \frac{-2}{\xs}+\frac{4}{y^2+\xg} \right]\frac{\delta\mu}{c_a}, \label{eqn:Sigma}
\end{align}
where $\xg = \dfrac{4\xs}{c_a}$ and $s_0^2$ is the spin-0 aether wave speed from (\ref{eqn:s02}).

This equation does not have an exact analytic solution; however, we can describe its behavior in  limiting regimes.  At short wavelengths, $s_0y\gg 1$, the equation takes the form of a Bessel equation with a time-dependent, decaying source.  Solutions thus behave like damped oscillators with a shifted equilibrium value.  In the opposite regime $s_0y\ll 1$, we obtain power law solutions describing growth on superhorizon scales.  The exponents are given by equation (\ref{eqns:exponent}) after setting $\Omega_\mu = 1$. 

\subsection{Constraints on $\Omega_\mu$}\label{sec:isw}
The superhorizon growth exponent $p_+$ from (\ref{eqns:exponent}) allows us to constrain the size of $\Omega_\mu$.  The quickest  constraint comes from requiring that the  potential $\gim$
remain linear throughout the epoch of matter domination.   This requirement is justified
since $M$ contributes to 
the gravitational potential; it is related to the usual Bardeen potentials 
$\Phi$ and $\Psi$ \cite{Bardeen1980,Peter2005} by
\begin{subequations}
\begin{align}
\Psi &= \giz-\gim+\hc V,\\
\Phi&= a^{-1}\left[a\left(\frac{\gim}{\hc}-V\right)\right]'.
\end{align}
\end{subequations}
Since these gravitational potentials are linear except on scales as small as stars 
or black holes, we require that $\gim$ should be linear as well.   Although
this leads to a rather modest constraint on $p_+$, it translates to a much more stringent constraint on $\Omega_\mu$, allowing us to conclude that $\mu$ cannot comprise a sizable fraction of the dark matter.

When the post-Newtonian constraints are satisfied by the parameter choice (\ref{eqn:ppnconstr}), the analysis of primordial perturbations for Einstein-aether theory \cite{Armendariz-Picon2010} reveals that both superhorizon modes found in section \ref{sec:scalarpert} should have comparable amplitudes of the order of $10^{-5}$ at the time of radiation-matter equality.   The superhorizon growing modes would have an amplitude today of
\begin{equation}
M_0 = M_\text{eq}(a_0/a_\text{eq})^{p_+/2}.
\end{equation}
With a redshift at equality of $z_\text{eq}\sim3500$ and initial amplitude $M_\text{eq}
\sim 10^{-5}$, $M_0$ will remain less than unity if 
\begin{equation}
p_+\lesssim 3.
\end{equation}

To translate this to a bound on $\Omega_\mu$, note that  
solar system tests of post-Newtonian parameters require $G_N$ and $G_\text{cos}$ to be very nearly equal \cite{Foster2006b,Will2006}. By examining the form of $p_+$ in (\ref{eqn:expG}), we see the $G_\text{cos}$ and $G_N$ terms cancel when the post-Newtonian constraints are satisfied, and the remaining term is proportional to $\Omega_\mu/c_a$.  Since $c_a$ is phenomenologically constrained to be small, $\Omega_\mu$ must be similarly small to produce an order unity growth exponent.  

More quantitatively, the preferred frame post-Newtonian parameters $\alpha_1$ and $\alpha_2$ can be set to zero in Einstein-aether theory by choosing \cite{Foster2006b}
\begin{equation} \label{eqn:ppnconstr}
-\frac{1}{c_\theta} = \frac{1}{c_a} = \frac12\left(\frac{1}{c_\sigma}+\frac{1}{c_\omega}\right).
\end{equation}
The \v{C}erenkov constraint imposes that both $c_\sigma$ and $c_\omega$ are positive, and observations of binary pulsar decay rates further imposes that \cite{Yagi2013}
\begin{subequations}
\begin{align}
c_\sigma\lesssim 3\times 10^{-2}, \\
c_\omega\lesssim 3\times 10^{-3},
\end{align}
\end{subequations}
which, combined with (\ref{eqn:ppnconstr}) gives
\begin{equation}
c_a\lesssim5\times 10^{-3}.
\end{equation}

Now, given a bound on the exponent $p_+\lesssim b$, and using the above restrictions on the aether coupling parameters, this translates to a bound on $\Omega_\mu$,
\begin{equation}
\Omega_\mu\lesssim \frac{b(b+5)}{12} c_a.
\end{equation}
Taking $b\sim3$, this leads to 
\begin{equation}
\Omega_\mu\lesssim 10^{-2}.
\end{equation}
Such a small value of $\Omega_\mu$ precludes it making any appreciable contribution to the dark matter in the universe.

The bounds on $p_+$ and $\Omega_\mu$ could be  improved  by considering the contribution of the growing modes to the integrated Sachs-Wolfe effect in the CMB.  The time varying gravitational potential $\Phi-\Psi$ during matter domination coming from the growing  modes  causes changes in photon wavelengths as they free stream from the surface of last scattering.  If the growth rate is too large, this will lead to an excess of power in the low $\ell$ multipoles 
of the CMB power spectrum.

Finally, we note that this constraint only applies for $c_a$ finite and nonzero.  In particular,
it does not apply to projectable \Horava gravity or the mimetic dark matter theory.  The 
growth exponent for those theories is obtained from the ponderable aether result by 
taking the limit $c_a\rightarrow\infty$ in (\ref{eqns:exponent}).  This results in $p_+=-2$
and $p_- = -3$, so that the isocurvature modes are decaying on superhorizon scales.  
These decaying modes would have little impact on CMB and large scale structure
observations.

\section{Discussion}\label{sec:conclusion}
This work has explored several consequences of treating the aether vector $u^a$ in a Lorentz-violating theory as nondynamical.  The resulting ponderable aether theory differs from dynamical Einstein-aether theory by the presence of source densities in the aether equation of motion.  The source densities take the form of a covector density $\tilde\mu_a$ that is conserved along flows of $u^a$.  They also have an interpretation in terms of the
violated initial value constraints of Einstein-aether theory (section \ref{sec:constraints}).

The most striking phenomenological feature of the source densities for ponderable
aethers  is that the scalar density $\tilde\mu$ can mimic the effects of cold dark matter in FLRW cosmologies (section \ref{sec:frw}).  However, beyond homogeneous, isotropic cosmolgy, $\tilde\mu$ differs from cold dark matter on several fronts.  First, as argued in section \ref{sec:cosmopheno}, the significant pressure in the aether fluid prevents gravitational collapse on scales shorter than the aether sound horizon, which is larger than the Hubble radius.  Thus it does not lead to the  structure formation observed in our universe.  Furthermore,  perturbations to the FLRW backgrounds, considered in section \ref{sec:scalarpert}, show that the ponderable aether produces a problematic growing mode on super-Hubble scales unless $\Omega_\mu$ is relatively small.  This precludes its interpretation as a significant component of the dark matter.  On smaller scales, $\mu$ does not allow for static, spherically symmetric solutions, so that it is unclear what the final state of gravitational collapse would be.

On the other hand, considerations of cosmological inflation provide a different perspective on this theory.  Under reasonable assumptions for the value of $\mu$ at the beginning of inflation, the requisite 60 e-foldings is more than enough to dilute it to a negligible value by the present era.  This is similar to
 the inflationary solution to the monopole
 problem \cite{Zeldovich1978, Preskill1979}, where
in our case $\mu$ is playing the role of an exactly stable massive particle.  
With $\mu$ diluted to nearly zero, the theory  essentially reduces to dynamical 
Einstien-aether theory.  Thus, a universe described by Einstein-aether theory could equally well be interpreted as a universe with a nondynamical, Lorentz-violating background structure, whose effects have been washed away by inflation. In this case, the line between the
diffeomorphism-invariant and noninvariant theories is not sharp.

Ponderable aethers are a vector field generalization of the mimetic dark matter construction
\cite{Chamseddine2013}, related to the model proposed in Ref. \cite{Barvinsky2014}.   
By taking the aether field to be the gradient of a scalar (as was considered in
\cite{Haghani2014a}), we obtain the IR limit of 
projectable \Horava gravity \cite{Blas2009, Jacobson2014a}.  The aether kinetic terms
then become the additional gradient terms that appear in extensions of the original
mimetic dark matter theory \cite{Mirzagholi2014, Capela2014}.  Note that these kinetic
terms would be expected from an effective field theory point of view, and are also 
necessary when the aether is nondynamical to avoid overly constrained gravitational 
dynamics \cite{Jacobson2014}.  We note that the constraint derived for ponderable
aethers in section \ref{sec:isw} does not apply for the mimetic and projectable \Horava 
theories, since these theories come from a $c_a\rightarrow\infty$ limit of the ponderable
aether theory, where there are no growing superhorizon modes.

Although the ponderable aether is highly constrained phenomenologically, it 
nevertheless presents some directions of theoretical interest for future work. 
The cosmological constraints derived in section \ref{sec:isw} were rough estimates, intended to only bound $\Omega_\mu$ enough to exclude its interpretation as the dark matter.  More detailed analysis of the cosmological implications could be performed, and likely could improve these bounds on $\Omega_\mu$ significantly.

The non-existence of static, spherically symmetric stars in this theory is a distinctive feature, and presents a puzzle regarding the final state of gravitational collapse.  This could be resolved by examining numerical simulations of gravitational collapse, along the lines of \cite{Garfinkle2007}.  Another possibility would be to look instead for stationary, axisymmetric solutions, which allow $\mu$ to orbit a central star or black hole without accreting.  
Another line of investigation  could focus on fixed threading theories, 
where $\mu=0$, but $\mu_a^\perp$ is nonzero.\footnote{Note that this is just a special 
case of the fixed aether theory where the initial value of $\mu$ is set to zero by hand.  
One can justify such a choice if the background structure is a fixed threading, as opposed
to a fixed aether \cite{Jacobson2014}.}  In that case, by choosing the aether vector to not align with the Killing vector, there may be solutions that generalize Einstein-aether black holes \cite{Eling2006a} with sourced spatial aether equations.

\begin{acknowledgments}
The author thanks N.\ Afshordi, L.\ Boyle, J.\ Sakstein,  K.\ Smith, 
and A. Vikman for helpful discussions,
and especially T.\ Jacobson for suggesting this topic as well as for extensive 
discussions and feedback. 
This research was supported in part by the 
National Science Foundation under grants No.\ PHY-0903572, PHY-1407744, 
and PHY11-25915, and by the Perimeter Institute for Theoretical Physics.  
Research at Perimeter Institute is supported by the Government of Canada through Industry 
Canada and by the Province of Ontario through the Ministry of Economic Development and
Innovation.
We also thank the UCSB Department of Physics for hospitality while this 
work was being completed.
\appendix

\section{Formulae for equations of motion} \label{sec:formulae}
The variational derivative of (\ref{eqn:aeaction}) with respect to the aether, $\eom_b = \dfrac{1}{\sqrt{-g}}\dfrac{\delta S_{\ae}}{\delta u^b}$, is given by
\begin{align}
\eom_b =& \frac{1}{8\pi G_0}\left[ \frac{c_\theta}{3} \nabla_b\theta + c_a(-a^2 u_b  +2 a^c\omega_{cb}+2\nabla^c(u_{[c}a_{b]})) \right. \nonumber \\
&\left. \vphantom{\frac{c_\theta}{3}} +  c_\omega(\nabla^c\omega_{cb}-a^c\omega_{cb}) + c_\sigma(\nabla^c\sigma_{cb}+a^c\sigma_{cb}) \right]  \label{eqn:aethereom}
\end{align}
The metric variation of (\ref{eqn:aeaction}) with $u^c$ held fixed, $\eom^{\ae}_{ab} =\dfrac{1}{\sqrt{-g}} \left.\dfrac{\delta S_{\ae}}{\delta g^{ab}}\right|_{u^c}$ takes the form
\begin{equation} \label{eqn:aemetricvar}
\eom^{\ae}_{ab} = \frac{-1}{16 \pi G_0}\left(G_{ab} + \frac{c_\theta}{3} \mathcal{X}_{ab}+c_a \mathcal{A}_{ab}+c_\sigma \mathcal{S}_{ab}+c_\omega \mathcal{V}_{ab} \right)
\end{equation}
where $G_{ab}$ is the Einstein tensor, and the other tensors appearing are
\begin{align}
\mathcal{X}_{ab}& = g_{ab}\left(\frac12\theta^2+\lie_u\theta \right)   \label{eqn:Xab} \\
\mathcal{A}_{ab}&=-\frac12 a^2g_{ab}+a_a a_b+2\nabla_c(u^c a_{(b}-a^c u_{(b}) u_{a)}  \\
\mathcal{S}_{ab} &= -\frac12 \sigma^2 g_{ab} -2\sigma\indices{_a^c}\sigma_{bc}+\frac13\theta\sigma_{ab} +\lie_u\sigma_{ab} \\
\mathcal{V}_{ab} &= -\frac12 g_{ab}\omega^2 +2\omega\indices{_a^c}\omega_{bc}+2(\nabla^c\omega_{c(a}  -a^c\omega_{c(a})u_{b)}
\end{align}

\section{Perturbative variables}\label{sec:pertvars}
In both the analysis of linearized perturbations to Minkowski space considered in section \ref{sec:linearwaves} and the analysis of cosmological perturbations in section \ref{sec:scalarpert}, it is convenient to use a scalar-vector-tensor (SVT) decomposition of the perturbative variables.  The reason is that the modes of different spins decouple in the linearized theory, and the SVT decomposition makes this explicit and also allows for the construction of gauge invariant variables.  Here, we describe the decomposition used throughout this paper, following closely the conventions of \cite{Peter2005}.  

Starting with cosmological perturbations, the flat background metric with conformal time is
\begin{equation}\label{eqn:flrwmetric}
ds^2 = a^2(\eta)(d\eta^2-\delta_{ij}dx^idx^j),
\end{equation}
and isotropy requires a background value for the aether to be
\begin{equation}
u^\alpha = (a^{-1},0,0,0).
\end{equation}
The perturbations to the metric are parameterized by
\begin{subequations}\label{eqns:gpert}
\begin{align}
g_{00} &=  a^2(1+2A) \label{eqn:g00pert} \\
g_{0i} &= -a^2(\partial_i B + B_i) \label{eqn:g0ipert} \\
g_{ij}&=-a^2(\delta_{ij}+h_{ij}) \label{eqn:gijpert}\\
h_{ij} &= 2C\delta_{ij}+2\partial_i\partial_j E + 2\partial_{(i}E_{j)} + 2 E_{ij},  \label{eqn:spatialpert}
\end{align}
\end{subequations}
and the aether perturbations by
\begin{subequations}\label{eqns:upert}
\begin{align}
u^0 &= a^{-1}(1-T) \label{eqn:u0}\\
u^i &= a^{-1}(\delta^{ij}\partial_j U+ U^i).
\end{align}
\end{subequations}
As is standard in the usual SVT decomposition, perturbed quantities carrying an index are divergenceless, i.e.\ $\partial_i B^i = \partial_i E^{ij}=0$, and the tensor perturbation is also traceless, $\delta^{ij}E_{ij}=0$.  Note also that we use $\delta_{ij}$ to raise and lower indices of perturbed quantities. 

From these we can construct gauge-invariant variables that appear in the linearized equations of motion.  The following gauge-invariant scalars are useful,
\begin{subequations} \label{eqns:gi}
\begin{align}
\gis &\equiv -C-\hc( B+U)\label{eqn:gis}\\
\gih&\equiv A+ (B+U)'+\hc(B+U) \label{eqn:gih}\\
\giv & \equiv E' + U\\
\gix&\equiv \hc\left(A-\left(\frac{C}{\hc}\right)'-C\right) \label{eqn:gix}\\
&=-\hc\left(\left(\frac{\gis}{\hc}\right)' +\gis+\gih \right), \label{eqn:gixalt}
\end{align}
\end{subequations}
where $'$ denotes differentiation with respect to $\eta$, and $\hc\equiv a'/a$.
A set of gauge-invariant vector variables are
\begin{subequations} \label{eqns:vgi}
\begin{align}
\giv_i &\equiv B_i + U_i \label{eqn:vgiv} \\
\giw_i&\equiv E_i'+U_i.
\end{align}
\end{subequations}
Also note that the tensor $E_{ij}$ is gauge-invariant.

The SVT decomposition of perturbations to Minkowski space can be obtained from these formulae by setting $a=1$ and $\hc=0$. In particular, the metric (\ref{eqn:flrwmetric}) becomes
\begin{equation}
ds^2 = dt^2-\delta_{ij}dx^idx^j,
\end{equation}
where we have renamed $\eta\rightarrow t$, and the background aether is just
\begin{equation}
u^\alpha = (1,0,0,0).
\end{equation}
For reference, we also list the relevant gauge invariant scalars:
\begin{subequations}\label{eqns:minkpert}
\begin{align}
\gis &= -C \\
\gih &= A + B' + U' \\
\giv&= E' +U.
\end{align}
\end{subequations}
The expressions for the vector gauge-invariants (\ref{eqns:vgi}) and the tensor gauge-invariant are unchanged.

\section{Perturbations in the aether gauge}\label{sec:pertcalcs}
The aether gauge choice greatly simplifies the computation of the quantities appearing in the aether action (\ref{eqn:aetheraction}) in perturbation theory. In this section, we describe the calculations that lead to the quadratic action for cosmological perturbation given in (\ref{eqn:quadaether}).  

The scalar part of the perturbed metric determinant is 
\begin{equation}
\sqrt{-g}_{(s)} =a^4\left(1+\frac12 h + h_2 \right),
\end{equation}
where $h\equiv \delta^{ij}h_{ij}=  6C+2\triangle E$, $h_2\equiv \frac18 h^2+\frac12\partial_i B \partial^i B-\frac14 h^{ij} h_{ij}$, and $\triangle = \delta^{ij}\partial_i\partial_j$ is the spatial Laplacian.  

This leads to the expansion
\begin{equation}
\theta = \frac1{\sqrt{-g}}\lie_u\sqrt{-g} = a^{-1}\left(3\hc +\frac12 h' +h'_2\right),
\end{equation}
which gives the quadratic order $c_\theta$ piece of (\ref{eqn:aetheraction}),
\begin{align}
\left(\sqrt{-g}\theta^2 \right)_{(2)}  
&=   a^2\left(\frac14(h')^2+ 9\hc^2h_2+6\hc h_2'\right), 
\end{align}
The last term can be integrated by parts in the action to give 
\begin{equation}
a^2\left(\frac14(h')^2-3(\hc^2+2\hc')h_2\right),\label{eqn:perttheta}
\end{equation}
and so in the special case of a matter-dominated background that we are considering, the second term drops out. 

Since the other parts of the Lagrangian (\ref{eqn:aetheraction}) vanish in the background, the acceleration and shear  only need to be computed to linear order, and the twist contains no scalar piece.  Also these quantities are orthogonal to $u^a$, which in the aether gauge means they have no $0$-component.  The acceleration is 
\begin{align}
a_\alpha &= \lie_u u_\alpha = -\delta^i_\alpha a^{-1}(a\partial_i B)'. \label{eqn:perta}
\end{align}
 The shear takes a simple form by first defining the spatial metric
\begin{equation}\label{eqn:project}
f_{ab}\equiv u_au_b - g_{ab},
\end{equation}
and then using the equation
\begin{align}
\sigma_{\alpha\beta}&=-\frac12 \lie_u f_{\alpha\beta}+\frac13\theta f_{\alpha \beta} \\
&=\frac12\delta_\alpha^i\delta_\beta^j\,a \left(h_{ij}-\frac13 h\delta_{ij}\right)'\nonumber \\
&=\delta_\alpha^i\delta_\beta^j\, a\left(\partial_i \partial_j E-\frac13\delta_{ij}\triangle E\right)'.\label{eqn:pertshear}
\end{align}

Contracting these quantities using the background metric yields the $c_\sigma$ and $c_a$ pieces of (\ref{eqn:quadaether}).

\end{acknowledgments}

\bibliographystyle{JHEP}
\bibliography{nondynamical_aether}

\providecommand{\href}[2]{#2}\begingroup\raggedright\begin{thebibliography}{10}

\bibitem{Jacobson2014}
T.~Jacobson and A.~J. Speranza, {\it Variations on an aethereal theme},  {\em
  arXiv eprints} (2014) [\href{http://arxiv.org/abs/1503.08911}{{\tt
  arXiv:1503.08911}}].

\bibitem{Boersma1995}
S.~Boersma and T.~Dray, {\it {Slicing, threading and parametric manifolds}},
  {\em Gen. Relativ. Gravit.} {\bf 27} (Mar., 1995) 319--339,
  [\href{http://arxiv.org/abs/gr-qc/9407020}{{\tt gr-qc/9407020}}].

\bibitem{Boersma1995a}
S.~Boersma and T.~Dray, {\it {Parametric manifolds. I. Extrinsic approach}},
  {\em J. Math. Phys.} {\bf 36} (July, 1995) 1378,
  [\href{http://arxiv.org/abs/gr-qc/9407011}{{\tt gr-qc/9407011}}].

\bibitem{Horava2009}
P.~Ho\v{r}ava, {\it {Quantum gravity at a Lifshitz point}},  {\em Phys. Rev. D}
  {\bf 79} (Apr., 2009) 084008, [\href{http://arxiv.org/abs/0901.3775}{{\tt
  arXiv:0901.3775}}].

\bibitem{Mukohyama2009}
S.~Mukohyama, {\it {Dark matter as integration constant in Ho\v{r}ava-Lifshitz
  gravity}},  {\em Phys. Rev. D} {\bf 80} (Sept., 2009) 064005,
  [\href{http://arxiv.org/abs/0905.3563}{{\tt arXiv:0905.3563}}].

\bibitem{Chamseddine2013}
A.~H. Chamseddine and V.~Mukhanov, {\it {Mimetic dark matter}},  {\em J. High
  Energy Phys.} {\bf 2013} (Nov., 2013) 135,
  [\href{http://arxiv.org/abs/1308.5410}{{\tt arXiv:1308.5410}}].

\bibitem{Golovnev2014}
A.~Golovnev, {\it {On the recently proposed mimetic Dark Matter}},  {\em Phys.
  Lett. B} {\bf 728} (Jan., 2014) 39--40,
  [\href{http://arxiv.org/abs/1310.2790}{{\tt arXiv:1310.2790}}].

\bibitem{Blas2009}
D.~Blas, O.~Pujol\`{a}s, and S.~Sibiryakov, {\it {On the extra mode and
  inconsistency of Ho\v{r}ava gravity}},  {\em J. High Energy Phys.} {\bf 2009}
  (Oct., 2009) 029, [\href{http://arxiv.org/abs/0906.3046}{{\tt
  arXiv:0906.3046}}].

\bibitem{Jacobson2014a}
T.~Jacobson and A.~J. Speranza, {\it {Comment on ``Scalar Einstein-Aether
  theory"}},  {\em arXiv eprints} (May, 2014)
  [\href{http://arxiv.org/abs/1405.6351}{{\tt arXiv:1405.6351}}].

\bibitem{Blas2010}
D.~Blas, O.~Pujol\`{a}s, and S.~Sibiryakov, {\it {Consistent Extension of
  Ho\v{r}ava Gravity}},  {\em Phys. Rev. Lett.} {\bf 104} (May, 2010) 181302,
  [\href{http://arxiv.org/abs/0909.3525}{{\tt arXiv:0909.3525}}].

\bibitem{Jacobson2013}
T.~Jacobson, {\it {Undoing the twist: the Ho\v{r}ava limit of
  Einstein-aether}},  {\em Phys. Rev. D} {\bf 89} (Apr., 2014) 081501,
  [\href{http://arxiv.org/abs/1310.5115}{{\tt arXiv:1310.5115}}].

\bibitem{Barvinsky2014}
A.~Barvinsky, {\it {Dark matter as a ghost free conformal extension of Einstein
  theory}},  {\em J. Cosmol. Astropart. P.} {\bf 2014} (Jan., 2014) 014,
  [\href{http://arxiv.org/abs/1311.3111}{{\tt arXiv:1311.3111}}].

\bibitem{Blas2012a}
D.~Blas, M.~M. Ivanov, and S.~Sibiryakov, {\it {Testing Lorentz invariance of
  dark matter}},  {\em J. Cosmol. Astropart. P.} {\bf 2012} (Oct., 2012) 057,
  [\href{http://arxiv.org/abs/1209.0464}{{\tt arXiv:1209.0464}}].

\bibitem{Jacobson2001}
T.~Jacobson and D.~Mattingly, {\it {Gravity with a dynamical preferred frame}},
   {\em Phys. Rev. D} {\bf 64} (June, 2001) 024028,
  [\href{http://arxiv.org/abs/gr-qc/0007031v4}{{\tt gr-qc/0007031v4}}].

\bibitem{Jacobson2008}
T.~Jacobson, {\it {Einstein-aether gravity: a status report}},  in {\em From
  Quantum to Emergent Gravity: Theory and Phenomenology}, (Trieste),
  Proceedings of Science, Jan., 2008.
\newblock \href{http://arxiv.org/abs/0801.1547}{{\tt arXiv:0801.1547}}.

\bibitem{Jacobson2011}
T.~Jacobson, {\it {Initial value constraints with tensor matter}},  {\em
  Classical Quant. Grav.} {\bf 28} (2011), no.~24
  [\href{http://arxiv.org/abs/1108.1496}{{\tt arXiv:1108.1496}}].

\bibitem{Seifert2007a}
M.~Seifert and R.~Wald, {\it {General variational principle for spherically
  symmetric perturbations in diffeomorphism covariant theories}},  {\em Phys.
  Rev. D} {\bf 75} (Apr., 2007) 084029,
  [\href{http://arxiv.org/abs/gr-qc/0612121v2}{{\tt gr-qc/0612121v2}}].

\bibitem{Andersson2007}
N.~Andersson and G.~L. Comer, {\it {Relativistic Fluid Dynamics: Physics for
  Many Different Scales}},  {\em Living Rev. Relativ.} {\bf 10} (2007)
  [\href{http://arxiv.org/abs/gr-qc/0605010}{{\tt gr-qc/0605010}}].

\bibitem{Dubovsky2006}
S.~Dubovsky, T.~Gr\'{e}goire, A.~Nicolis, and R.~Rattazzi, {\it {Null energy
  condition and superluminal propagation}},  {\em J. High Energy Phys.} {\bf
  2006} (2006), no.~03 025, [\href{http://arxiv.org/abs/hep-th/0512260v2}{{\tt
  hep-th/0512260v2}}].

\bibitem{Dubovsky2012}
S.~Dubovsky, L.~Hui, A.~Nicolis, and D.~T. Son, {\it {Effective field theory
  for hydrodynamics: Thermodynamics, and the derivative expansion}},  {\em
  Phys. Rev. D} {\bf 85} (Apr., 2012) 085029,
  [\href{http://arxiv.org/abs/1107.0731}{{\tt arXiv:1107.0731}}].

\bibitem{Jacobson2004}
T.~Jacobson and D.~Mattingly, {\it {Einstein-aether waves}},  {\em Phys. Rev.
  D} {\bf 70} (July, 2004) 024003,
  [\href{http://arxiv.org/abs/gr-qc/0402005}{{\tt gr-qc/0402005}}].

\bibitem{Foster2006}
B.~Foster, {\it {Radiation damping in Einstein-aether theory}},  {\em Phys.
  Rev. D} {\bf 73} (May, 2006) 104012,
  [\href{http://arxiv.org/abs/gr-qc/0602004}{{\tt gr-qc/0602004}}].

\bibitem{Carroll2004}
S.~Carroll and E.~Lim, {\it {Lorentz-violating vector fields slow the universe
  down}},  {\em Phys. Rev. D} {\bf 70} (Dec., 2004) 123525,
  [\href{http://arxiv.org/abs/hep-th/0407149}{{\tt hep-th/0407149}}].

\bibitem{Eling2006b}
C.~Eling and T.~Jacobson, {\it {Spherical Solutions in Einstein-Aether Theory:
  Static Aether and Stars}},  {\em Classical Quant. Grav.} {\bf 23} (Mar.,
  2006) 5625, [\href{http://arxiv.org/abs/gr-qc/0603058}{{\tt gr-qc/0603058}}].

\bibitem{Izumi2010}
K.~Izumi and S.~Mukohyama, {\it {Stellar center is dynamical in
  Ho\v{r}ava-Lifshitz gravity}},  {\em Phys. Rev. D} {\bf 81} (Feb., 2010)
  044008, [\href{http://arxiv.org/abs/0911.1814}{{\tt arXiv:0911.1814}}].

\bibitem{Eling2006a}
C.~Eling and T.~Jacobson, {\it {Black holes in Einstein-aether theory}},  {\em
  Classical Quant. Grav.} {\bf 23} (Sept., 2006) 5643--5660,
  [\href{http://arxiv.org/abs/gr-qc/0604088}{{\tt gr-qc/0604088}}].

\bibitem{Mirzagholi2014}
L.~Mirzagholi and A.~Vikman, {\it {Imperfect Dark Matter}},  {\em arXiv
  eprints} (Dec., 2014) [\href{http://arxiv.org/abs/1412.7136}{{\tt
  arXiv:1412.7136}}].

\bibitem{Elliott2005}
J.~W. Elliott, G.~D. Moore, and H.~Stoica, {\it {Constraining the New Aether:
  gravitational Cherenkov radiation}},  {\em J. High Energy Phys.} {\bf 2005}
  (Aug., 2005) 066, [\href{http://arxiv.org/abs/hep-ph/0505211v2}{{\tt
  hep-ph/0505211v2}}].

\bibitem{Lim2005}
E.~Lim, {\it {Can we see Lorentz-violating vector fields in the CMB?}},  {\em
  Phys. Rev. D} {\bf 71} (Mar., 2005) 063504,
  [\href{http://arxiv.org/abs/astro-ph/0407437v2}{{\tt astro-ph/0407437v2}}].

\bibitem{Li2008}
B.~Li, D.~Mota, and J.~Barrow, {\it {Detecting a Lorentz-violating field in
  cosmology}},  {\em Phys. Rev. D} {\bf 77} (Jan., 2008) 024032,
  [\href{http://arxiv.org/abs/0709.4581}{{\tt arXiv:0709.4581}}].

\bibitem{Armendariz-Picon2010}
C.~Armendariz-Picon, N.~F. Sierra, and J.~Garriga, {\it {Primordial
  perturbations in Einstein-Aether and BPSH theories}},  {\em J. Cosmol.
  Astropart. P.} {\bf 2010} (July, 2010) 010,
  [\href{http://arxiv.org/abs/1003.1283}{{\tt arXiv:1003.1283}}].

\bibitem{Sierra2011}
N.~F. Sierra, {\em {Cosmological Perturbations in Einstein-Aether Theories}}.
\newblock Ph.d. thesis, University of Barcelona, 2011.

\bibitem{Lim2010}
E.~A. Lim, I.~Sawicki, and A.~Vikman, {\it {Dust of dark energy}},  {\em J.
  Cosmol. Astropart. P.} {\bf 2010} (May, 2010) 012,
  [\href{http://arxiv.org/abs/1003.5751}{{\tt arXiv:1003.5751}}].

\bibitem{Zlosnik2008}
T.~Zlosnik, P.~Ferreira, and G.~Starkman, {\it {Growth of structure in theories
  with a dynamical preferred frame}},  {\em Phys. Rev. D} {\bf 77} (Apr., 2008)
  084010, [\href{http://arxiv.org/abs/0711.0520}{{\tt arXiv:0711.0520}}].

\bibitem{Kobayashi2010a}
T.~Kobayashi, Y.~Urakawa, and M.~Yamaguchi, {\it {Cosmological perturbations in
  a healthy extension of Horava gravity}},  {\em J. Cosmol. Astropart. P.} {\bf
  2010} (Feb., 2010) 10, [\href{http://arxiv.org/abs/1002.3101}{{\tt
  arXiv:1002.3101}}].

\bibitem{Papazoglou2010}
A.~Papazoglou and T.~P. Sotiriou, {\it {Strong coupling in extended
  Ho\v{r}ava–Lifshitz gravity}},  {\em Phys. Lett. B} {\bf 685} (Mar., 2010)
  197--200, [\href{http://arxiv.org/abs/0911.1299}{{\tt arXiv:0911.1299}}].

\bibitem{Blas2010a}
D.~Blas, O.~Pujol\`{a}s, and S.~Sibiryakov, {\it {Comment on “Strong coupling
  in extended Ho\v{r}ava–Lifshitz gravity” [Phys. Lett. B 685 (2010)
  197]}},  {\em Phys. Lett. B} {\bf 688} (May, 2010) 350--355,
  [\href{http://arxiv.org/abs/0912.0550}{{\tt arXiv:0912.0550}}].

\bibitem{Withers2009}
B.~Withers, {\it {Einstein-aether as a quantum effective field theory}},  {\em
  Classical Quant. Grav.} {\bf 26} (Nov., 2009) 225009,
  [\href{http://arxiv.org/abs/0905.2446}{{\tt arXiv:0905.2446}}].

\bibitem{Bardeen1980}
J.~Bardeen, {\it {Gauge-invariant cosmological perturbations}},  {\em Phys.
  Rev. D} {\bf 22} (Oct., 1980) 1882--1905.

\bibitem{Peter2005}
P.~Peter and J.-P. Uzan, {\em Primordial Cosmology}.
\newblock Oxford University Press, 2005.

\bibitem{Foster2006b}
B.~Foster and T.~Jacobson, {\it {Post-Newtonian parameters and constraints on
  Einstein-aether theory}},  {\em Phys. Rev. D} {\bf 73} (Mar., 2006) 064015,
  [\href{http://arxiv.org/abs/gr-qc/0509083v2}{{\tt gr-qc/0509083v2}}].

\bibitem{Will2006}
C.~M. Will, {\it {The Confrontation between General Relativity and
  Experiment}},  {\em Living Rev. Relativ.} {\bf 9} (2006)
  [\href{http://arxiv.org/abs/gr-qc/0510072}{{\tt gr-qc/0510072}}].

\bibitem{Yagi2013}
K.~Yagi, D.~Blas, N.~Yunes, and E.~Barausse, {\it {Strong Binary Pulsar
  Constraints on Lorentz Violation in Gravity}},  {\em Phys. Rev. Lett.} {\bf
  112} (July, 2014) 161101, [\href{http://arxiv.org/abs/1307.6219}{{\tt
  arXiv:1307.6219}}].

\bibitem{Zeldovich1978}
Y.~Zeldovich and M.~Khlopov, {\it {On the concentration of relic magnetic
  monopoles in the universe}},  {\em Phys. Lett. B} {\bf 79} (Nov., 1978)
  239--241.

\bibitem{Preskill1979}
J.~Preskill, {\it {Cosmological Production of Superheavy Magnetic Monopoles}},
  {\em Phys. Rev. Lett.} {\bf 43} (Nov., 1979) 1365--1368.

\bibitem{Haghani2014a}
Z.~Haghani, T.~Harko, H.~R. Sepangi, and S.~Shahidi, {\it {The Scalar
  Einstein-Aether theory}},  {\em arXiv eprints} (Apr., 2014)
  [\href{http://arxiv.org/abs/1404.7689}{{\tt arXiv:1404.7689}}].

\bibitem{Capela2014}
F.~Capela and S.~Ramazanov, {\it {Modified Dust and the Small Scale Crisis in
  CDM}},  {\em arXiv eprints} (Dec., 2014)
  [\href{http://arxiv.org/abs/1412.2051}{{\tt arXiv:1412.2051}}].

\bibitem{Garfinkle2007}
D.~Garfinkle, C.~Eling, and T.~Jacobson, {\it {Numerical simulations of
  gravitational collapse in Einstein-aether theory}},  {\em Phys. Rev. D} {\bf
  76} (July, 2007) 024003, [\href{http://arxiv.org/abs/gr-qc/0703093v2}{{\tt
  gr-qc/0703093v2}}].

\end{thebibliography}\endgroup

\end{document}